\documentclass[conference]{IEEEtran}
\IEEEoverridecommandlockouts
\usepackage{cite}
\usepackage{amsmath,amssymb,amsfonts}
\usepackage{graphicx}
\usepackage{textcomp}
\usepackage{xcolor}
\def\BibTeX{{\rm B\kern-.05em{\sc i\kern-.025em b}\kern-.08em
    T\kern-.1667em\lower.7ex\hbox{E}\kern-.125emX}}
  
\usepackage{amssymb}
\usepackage{gensymb}
\usepackage{graphicx}
\graphicspath{ {images/} }
\usepackage{amsmath}
\usepackage{float}
\usepackage{algorithm}
\usepackage{algpseudocode}
\usepackage{makecell}

\begin{document}

\title{Rate-Splitting Multiple Access for Multigroup Multicast Cellular and Satellite Communications: PHY Layer Design and Link-Level Simulations 
}

\author{\IEEEauthorblockN{Longfei Yin, Onur Dizdar and Bruno Clerckx}
\IEEEauthorblockA{Department of Electrical and Electronic Engineering, Imperial College London, United Kingdom\\
$\mathrm{Email:}\left \{\mathrm{longfei.yin17, 
o.dizdar, b.clerckx} \right \}\mathrm{@imperial.ac.uk}$
\thanks{This work was partially supported by the U.K. Engineering and Physical Sciences Research Council (EPSRC) under EP/R511547/1.}
}
}

\maketitle

\begin{abstract}
Rate-splitting multiple access (RSMA), relying on linearly precoded rate-splitting (RS) at the transmitter and successive interference cancellation (SIC) at the receivers 
has emerged as a powerful and flexible multiple access strategy for downlink multi-user multi-antenna systems.
Through message splitting and the transmission of both common and private messages, RSMA has been demonstrated to be a robust interference management strategy which enables partially decoding interference and partially treating interference as noise.
In this work, we consider the application of RSMA in a multigroup multicast scenario, where each message is intended to a group of users.
By leveraging the recent results on the max-min fair (MMF) optimization problem of RSMA-based multigroup multicast beamforming with imperfect channel state information at the transmitter (CSIT), we investigate the design of the physical (PHY) layer including finite length polar coding, finite alphabet modulation, adaptive modulation and coding (AMC) algorithm, and SIC receivers, etc. 
Link-level simulation (LLS) results 
verify the superiority of RSMA-based multigroup multicast transmission compared with space-division multiple access (SDMA)-based strategy in both cellular systems and multibeam satellite systems.

\end{abstract}


\section{Introduction}
With the explosion growth of data traffic and the rapid development of Internet of Things in 5G, demands for wireless communications such as high throughput, massive
connectivity of devices, content-centric services and heterogeneity of service types are continuously rising.
Owing to such requirements, multi-user (MU) multiple-input single/multiple-output (MIMO) and multiple access strategies have received considerable attention in both in academia and industry.

Due to the promising performance in a wide range of network loads (underloaded or overloaded regimes), channel disparity, channel orthogonality and channel state information at the transmitter (CSIT) imperfectness, rate-splitting multiple access (RSMA) 
has recently
emerged as a powerful non-orthogonal transmission and
robust interference management strategy for multi-antenna
wireless networks.
In \cite{8907421}, RSMA is shown analytically to generalize and encompass four seemingly different strategies, namely space division multiple cccess (SDMA) based
on linear precoding, orthogonal multiple cccess (OMA), power-domain nonorthogonal multiple access (NOMA) based on linearly precoded superposition coding with successive interference cancellation (SIC), and physical-layer multicasting.
The key behind the flexibility and robust manner of RSMA is to split each message into a common part and a private part. 
All common parts are jointly encoded into a common stream to be decoded by all users, while the private parts are individually encoded into private streams. 
After the the common stream is decoded and subtracted by SIC, each user then decodes its desired private stream and treats the remaining interference as noise. 
Such framework enables RSMA to partially decode interference and partially treat interference as noise.
The benefits achieved by RSMA over conventional strategies have been demonstrated in
various multi-antenna scenarios, namely multiuser unicast with perfect CSIT \cite{mao2018rate, ahmad2019interferenc, zhang2019cooperative } and imperfect CSIT \cite{7555358, 7972900, 7513415, 7152864, 7434643,9158344 }, multigroup multicast \cite{8019852,8926413,9130871,8445878,9257433}, and superimposed unicast and multicast \cite{mao2019rate}, etc. 

Differing from the aforementioned works which investigate the performance of RSMA with assumptions of Gaussian inputs and infinite block length, this work studies
the RSMA physical (PHY) layer architecture for multigroup multicast with finite length polar coding, finite alphabet modulation, adaptive modulation and coding algorithm (AMC) and SIC receivers, etc.
In \cite{9217103}, the uncoded link-level performance of RSMA-based MU-MISO systems is investigated.
With channel coding taken into consideration,
\cite{9217326} designs the basic transmitter and receiver architecture for RMSA in a MISO broadcast channel (BC) with two single-antenna users.
In this paper, by leveraging the same architecture as \cite{9217326} and the recent results on optimized RSMA-based multigroup multicast beamforming \cite{9257433}, 
link-level simulations (LLS) are conducted to show explicit throughput gain of RSMA over SDMA for multigroup multicasting in both cellular and multibeam satellite systems.

The rest of this paper is organized as follows: the system model and RSMA PHY-layer architecture for multigroup multicast is
described in Section II. 
Link-level simulation results in both cellular and multi beam satellite systems are illustrated in Section III.
Finally, section IV concludes the paper.

\textit{Notations}: 
Boldface, lowercase and standard letters denote matrices, column vectors, and scalars, respectively. 
$\mathbb{R}$ and $\mathbb{C}$ represent the real and complex domains. 
The real part of a complex number $x$ is given by $\mathcal{R}\left (x  \right )$.
$\left ( \cdot  \right )^{T}$ and $\left ( \cdot  \right )^{H}$ denote the transpose and the Hermitian transpose respectively. 
$\left | \cdot  \right |$ and $\left \| \cdot  \right \|$ denote the absolute value of a scalar or the cardinality of a set and the Euclidean norm.
$\mathbf{I}$ denotes the identity matrix. 
$\otimes $ denotes the Kronecker product and $\mathbf{D}^{\otimes }$ denotes Kronecker power of a matrix.

\section{RSMA PHY-Layer Architecture}

We consider a multigroup multicast MISO downlink system, comprising a transmitter with $N_{t}$ transmit antennas and $K$ single-antenna users denoted by $\mathcal{K}=\left \{ 1, \cdots, K \right \}$.
The set of $M$ multicast groups is denoted by $\mathcal{M}=\left \{ 1, \cdots, M \right \}$.
Messages are independent amongst different groups.
Each user $k \in \mathcal{K}$ belongs to only one group $\mathcal{G}_{m}$, where $m \in \mathcal{M}$.
The size of each group is denoted by $G_{m}=\left |\mathcal{G}_{m}   \right |$.
We have $\mathcal{G}_{i} \ \cap\  \mathcal{G}_{j} = \varnothing $ for all $i,j \in \mathcal{M}$, $i \neq j$ and $\cup _{m \in \mathcal{M}} \mathcal{G}_{m} = \mathcal{K}$.
By applying RSMA strategy,
each message intended for group-$m$, $W_{m}$, for all $ m \in \mathcal{M}$, is split into a common part $W_{m,c}$ and a private part $W_{m,p}$.
All common parts of the messages are combined into a common message $W_{c}$.
The common message $W_{c}$ and all private parts of the messages are respectively encoded into $s_{c}$ and $s_{1},\cdots, s_{M}$.
Thus, the vector of symbol streams is $\mathbf{s}=\left [ s_{c} ,s_{1},\cdots ,s_{M}\right ]^{T}\in \mathbb{C}^{\left (M+1 \right ) \times 1}$, where $\mathbb{E}\big \{ \mathbf{s} \mathbf{s}^{H} \big \}= \mathbf{I}$. 
The linear precoding vectors are denoted by $\mathbf{p}_{c},\mathbf{p}_{1 } , \cdots, \mathbf{p}_{M }\in \mathbb{C}^{N_{t} \times 1}$.
$\mathbf{P}=\left [ \mathbf{p}_{c},\mathbf{p}_{1},\cdots \mathbf{p}_{M} \right ]\in \mathbb{C}^{N_{t}\times \left (M+1 \right )}$ is the precoding matrix.
Thus, the transmit signal writes as
\begin{equation}
\mathbf{x} = \mathbf{P}\mathbf{s} = \mathbf{\mathbf{p}}_{c}s_{c} + \sum_{m=1}^{M}\mathbf{\mathbf{p}}_{m}s_{m}.
\end{equation}
When a sum transmit power constraint is taken into account, we have $\mathrm{tr}\left ( \mathbf{P}\mathbf{P}^{H} \right ) = \left \| \mathbf{p}_{c} \right \|^{2} + \sum _{m \in \mathcal{M}} \left \| \mathbf{p}_{m} \right \|^{2} \leq P_{t}$, where $P_{t}$ is the sum transmit power limit.
For per-antenna transmit power constraints, we have $\left ( \mathbf{P}\mathbf{P}^{H}   \right )_{n,n} \leq P_{n}, \ \forall n =1, \cdots,N_{t}$, where
$P_{n}$ is the power limit for antenna-$n$. 
We can rewrite the transmit power constraints generally as
\begin{equation}
\mathbf{p}_{c}^{H}\mathbf{D}_{l}\mathbf{p}_{c}+\sum_{m=1}^{M}\mathbf{p}_{m}^{H}\mathbf{D}_{l}\mathbf{p}_{m}\leq P_{l}, \ L \in \left \{ 1,\cdots,L \right \},
\end{equation}
where $L$ is defined as the number of transmit power constraints.
$\mathbf{D}_{l}$ is a diagonal shaping matrix that changes according to the system demands.
When the focus is on a sum transmit power constraint, let $L =1,\ \mathbf{D}_{l}=\mathbf{I}$ and  $P_{l} = P_{t}$.
In some practical implementations using per-antenna transmit power constraints (e.g., satellite communications), let $L = N_{t}$ and let matrix ${D}_{l}$ be a zero matrix except its $l$-th diagonal element equaling to $1$. 
The received signal at user-$k$ is written as
\begin{equation} 
    y_{k} = \mathbf{h}_{k}^{H}\mathbf{x} + n_{k},\ \forall k\in \mathcal{K},
\end{equation}
where $\mathbf{h}_{k}\in \mathbb{C}^{N_{t} \times 1}$ represents the channel vector between the transmitter and user-$k$.
The composite channel matrix is given by $\mathbf{H} = \left [ \mathbf{h}_{1}, \cdots,\mathbf{h}_{K} \right ]$, and 
$n_{k}\sim \mathcal{CN}\big ( 0,1 \big )$ is the Additive White Gaussian Noise (AWGN).
For convenience, we define $\mu:\mathcal{K}\rightarrow \mathcal{M}$ as mapping a user to its corresponding group, i.e., $\mu\left ( k \right ) = m, \ \forall k \in \mathcal{G}_{m}$.
On the receiver sides, each user at first detects the commons stream and treats all private streams as noise.
The common message estimate $\widehat{W}_{c}$ is obtained.
After reconstructing and subtracting the common stream through SIC,
each user then detects its private message $\widehat{W}_{p,\mu\left ( k \right )}$. 
$\widehat{W}_{\mu\left ( k \right )}$ is the message estimate of user-$k$, and can be obtained by combining $\widehat{W}_{c,\mu\left ( k \right )}$ and $\widehat{W}_{p,\mu\left ( k \right )}$.


\begin{figure*} 
    \centering
    \includegraphics[width=1\textwidth]{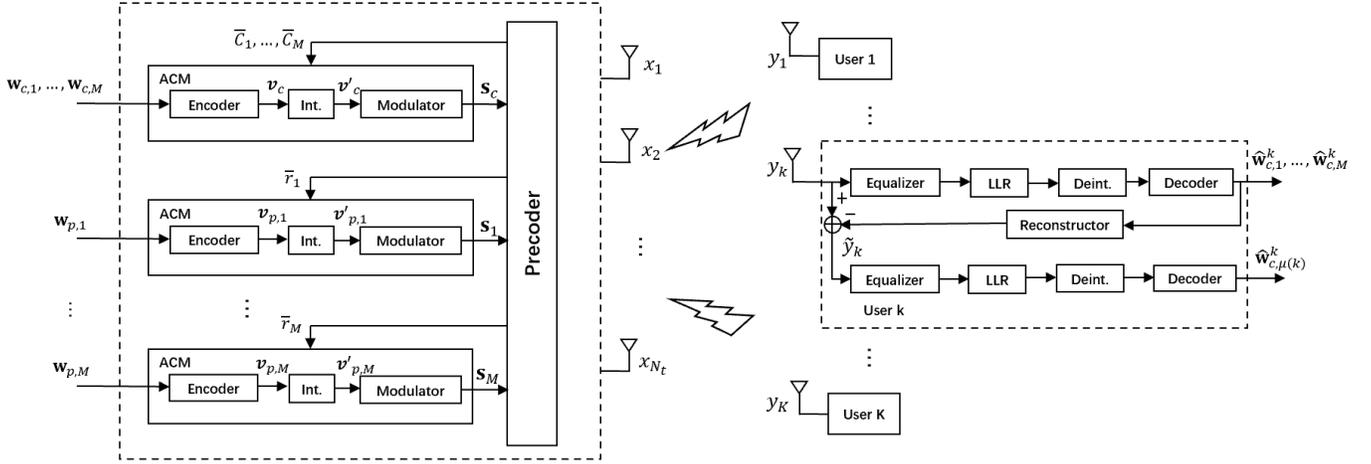}
    \caption{RSMA transmitter and receiver architecture.}
    \label{fig:Fig1}
\end{figure*}

The Signal-to-Interference-plus-Noise Ratio (SINR) of decoding $s_{c}$ at user-$k$ is
\begin{equation}
\gamma_{c,k}=\frac{\left | \mathbf{h}_{k}^{H} \mathbf{p}_{c} \right |^{2}}{ \left |\mathbf{h}_{k}^{H} \mathbf{p}_{\mu\left ( k \right )}   \right |^{2}+ \sum^{M}_{j=1, j\neq \mu\left ( k \right )}\left |\mathbf{h}_{k}^{H} \mathbf{p}_{j}   \right |^{2} +\sigma _{n}^{2}\ }.
\end{equation}
Its corresponding rate writes as $R_{c,k} =\log_{2} \left ( 1+ \gamma_{c,k} \right )$. To guarantee that each user is capable of decoding $s_{c}$, the common rate $R_{c}$ at which $s_{c}$ is communicated is defined as
\begin{equation}
R_{c}\triangleq \min _{k\in \mathcal{K}} R_{c,k}.
\end{equation}
Note that $s_{c}$ is shared among groups such that $R_{c} \triangleq \sum_{m=1}^{M}C_{m}$, where $C_{m}$ corresponds to group-$m$'s portion of common rate. 
After the common stream $s_{c}$ is decoded and removed through SIC, each user then decodes its desired private stream by treating all the other interference streams as noise.
The SINR of decoding $s_{\mu\left ( k \right )}$ at user-$k$ is given by
\begin{equation}
\gamma_{k}=\frac{\left | \mathbf{h}_{k}^{H} \mathbf{p}_{\mu\left ( k \right )} \right |^{2}}{ \sum^{M}_{j=1, j\neq \mu\left ( k \right )}\left |\mathbf{h}_{k}^{H} \mathbf{p}_{j}   \right |^{2} +\sigma _{n}^{2}\ }.
\end{equation}
The corresponding rate is written as $R_{k} =\log_{2} \left ( 1+ \gamma_{k} \right )$.
In terms of group-$m$, the multicast information $s_{m}$ should be decoded by all users in $\mathcal{G}_{m}$. Thus, the shared information rate $r_{m}$ is determined by the weakest user in $\mathcal{G}_{m}$ and defined by
\begin{equation}
r_{m}\triangleq \min _{i\in \mathcal{G}_{m} }  R_{i}.
\end{equation}
The $m$-th group-rate is the sum of $C_{m}$ and $r_{m}$, and writes as
\begin{equation}
r_{g,m}^{RS} = C_{m} + r_{m} = C_{m} + \min _{i\in \mathcal{G}_{m} }  R_{i}.
\end{equation}

We consider the imperfect CSIT model in \cite{7555358}, where the channel matrix is given by $\mathbf{H} = \widehat{\mathbf{H}} + \widetilde{\mathbf{H}}$.
Specifically, $\widehat{\mathbf{H}}$ is the channel estimate at the transmitter.
$\widetilde{\mathbf{H}} = \big [\widetilde{ \mathbf{h}}_{1}, \cdots,\widetilde{\mathbf{h}}_{K} \big ]$ represents the CSIT error with entries independent and identically drawn from $\mathcal{CN}\big ( 0,\sigma _{e}^{2} \big )$.
$\sigma _{e}^{2}= P^{- \alpha}$ is the CSIT error variance, and $\alpha \in \left [ 0,1 \right ]$ is the CSIT scaling factor.
In this paper, we use the beamforming scheme in \cite{9257433}, where all optimized precoders are obtained by solving an ergodic max min fair (MMF) group-rate problem under imperfect CSIT and assumptions of Gaussian inputs and infinite block length. 
Details of the algorithm and solutions can be found in \cite{9257433, 9145200}. 
The optimized precoders are utilized to calculate the average rates (ARs) of the common and private streams so as to determine appropriate modulation and coding in the adaptive modulation and coding (AMC) algorithm elaborated in the following subsection.

In this work, we investigate the PHY layer design and link-level performance of RSMA in a multigroup multicast system.
The RSMA transmitter and receiver architecture for multigroup multicast is depicted in Fig. 1.
We use finite alphabet modulation symbols carrying codewords from finite length polar code codebooks as channel inputs. 
The overall framework follows the architecture in \cite{9217326}, where a 2-user MISO BC system is considered.
Detailed explanations of each module are as follows.

\subsection{Encoder}
From Fig. 1,
$\mathbf{w}_{c,1},\cdots,\mathbf{w}_{c,M}$ represent all common parts of the group messages, which are bit vectors of length $K_{c,1},\cdots,K_{c,M}$. 
All private parts of the group messages are denoted by
$\mathbf{w}_{p,1},\cdots,\mathbf{w}_{p,M}$, which are bit vectors of length $K_{p,1},\cdots,K_{p,M}$.
Through the encoder, all common parts $\mathbf{w}_{c,1},\cdots,\mathbf{w}_{c,M}$ are jointly encoded into a common codeword $\mathbf{\nu}_{c}$ of code block length $N_{c}$, 
while the private parts
$\mathbf{w}_{p,1},\cdots,\mathbf{w}_{p,M}$ are encoded individually into private codewords $\mathbf{\nu}_{p,1},\cdots,\mathbf{\nu}_{p,M}$.
The code block lengths are respectively $N_{p,1},\cdots,N_{p,M}$.
We consider polar coding for the channel coding process. 
The block length of a conventional polar code is expressed as 
$N = 2^{n}$, where $n$ is a positive integer.
The polar encoding operation can be written as $\mathbf{\nu}= \mathbf{u}\mathbf{G}_{N}$, where $\mathbf{G}_{N}= \mathbf{B}_{N} 
\begin{bmatrix}
1 & 0\\ 
1 & 1
\end{bmatrix}^{\otimes n }$.
$\mathbf{B}_{N} $ is the bit-reversal matrix and
$\otimes n $ represents the $n$-fold Kronecker product. 
$\mathbf{u}$ denotes the length-$N$ uncoded bit vector input to the encoder which
consists of $K$ information bits and $N-K$ frozen bits.
Let $\mathcal{A} \in \left \{ 1, \cdots, N \right \}$ be the set of positions of the information bits, and
$\mathcal{A}^{c}$ be the set of positions of the frozen bits.
Therefore, we have $\mathcal{A} \cap \mathcal{A}^{c} = \phi $ and 
$\mathcal{A} \cup  \mathcal{A}^{c} = \left \{ 1,\cdots, N \right \}$.
Specifically, 
we can construct the private uncoded bit vectors $\mathbf{u}_{p,1},\cdots, \mathbf{u}_{p,M}$ by setting $\mathbf{u}_{p,m,\mathcal{A}_{m}}=\mathbf{w}_{p,m}, \forall m \in \mathcal{M}$.
The sets $\mathcal{A}_{p,1},\cdots, \mathcal{A}_{p,M}$ contain information bit indices of the private messages.
To jointly encode the common information bit vectors, $\mathbf{w}_{c,1},\cdots,\mathbf{w}_{c,M}$ are at first appended into $\mathbf{w}_{c}=\left [\mathbf{w}_{c,1},\cdots,\mathbf{w}_{c,M}  \right ]$.
Then, the common uncoded bit vector $\mathbf{u}_{c}$ is constructed by setting 
$\mathbf{u}_{c,\mathcal{A}_{c}}=\mathbf{w}_{c}$, where the set $\mathcal{A}_{c}$ collects information bit indices of the common message.
Values of all frozen bits are fixed and known by both encoder and decoder.
In addition, Cyclic Redundancy Check (CRC) codes are used as outer codes for all private and common messages, in order to enhance the error performance of the polar codes \cite{7055304}.
After obtaining the codewords 
$\mathbf{\nu}_{c}$ and 
$\mathbf{\nu}_{1},\cdots,\mathbf{\nu}_{M}$, interleavers are adopted before modulation.

\subsection{Modulator}
The interleavered bit vectors $\mathbf{\nu'}_{c}, \mathbf{\nu'}_{1},\cdots,\mathbf{\nu'}_{M}$
are respectively modulated into a common stream $\mathbf{s}_{c}$ and private streams $\mathbf{s}_{1},\cdots,\mathbf{s}_{M}$.
For a given modulation scheme with alphabet $\mathcal{M}$
and modulation order $\left |\mathcal{M}  \right |=2^{m}$. 
The interleavered bits $\left ( \nu' _{mi+1} , \cdots, \nu' _{mi+m}\right )$, $i \in \left \{ 0,1,\cdots, \frac{N}{m}-1 \right \}$ are mapped to a constellation signal $s \in \mathcal{M}$ according to the Gray labeling.
If a stream $\mathbf{s}$ is of length $S$, its corresponding code block length is $N = mS$.
\subsection{AMC Algorithm}
Appropriate modulation schemes and coding parameters are determined by the adaptive modulation and coding (AMC) algorithm to maximize the system throughput level depending on the channel characteristics.
The algorithm uses the average rates (ARs) $\overline{R}_{c}$ and $\overline{r}_{1},\cdots, \overline{r}_{M}$ obtained from the MMF optimization problem with imperfect CSIT in \cite{9257433} as link quality metrics.
The ARs of the common and private streams are actually calculated based on the optimized precoders by taking an average over $1000$ channel realizations due to the effects of imperfect CSIT.
Details of the algorithm and solutions can be found in \cite{9257433, 9145200}.
According to each given AR, we at first determine a corresponding modulation scheme from a modulation alphabet set $\mathcal{Q}$.
Here, we consider Quadrature Amplitude Modulation (QAM) schemes including $4$-QAM, $16$-QAM, $64$-QAM and $256$-QAM.
The set of feasible modulation schemes for a given AR $\overline{R}_{l} \in \left \{ \overline{R}_{c},\overline{r}_{1},\cdots, \overline{r}_{M} \right \}$
is given by
\begin{equation}
\mathcal{Q}\left (\overline{R}_{l},\beta  \right )=\left \{ \mathcal{M}:\mathrm{log}_{2}\left | \mathcal{M} \right |
\geq \mathrm{min}\left ( \frac{\overline{R}_{l}}{\beta },m' \right ),\mathcal{M}\in \mathcal{Q}\right \}.
\end{equation}
where $\beta$ is the maximum code rate indicating the proportion of information.
$m'$ is the logarithm of the highest modulation order, i.e., $m'=8$ for $256$-QAM in this paper.
For all $\overline{R}_{l} \in \left \{ \overline{R}_{c},\overline{r}_{1},\cdots, \overline{r}_{M} \right \}$, the modulation alphabets of the common and private streams are determined by
\begin{equation}
    \mathcal{M}_{l} = \mathrm{argmin}_{\mathcal{M} \in \mathcal{Q}\left (\overline{R}_{l},\beta  \right )}\left | \mathcal{M} \right |,\ \forall 
    l \in \left \{ c,1,\cdots,M \right \}.
\end{equation}
Thus, when all the streams are of length $S$, the code block lengths and code rates are respectively calculated as
\begin{equation}
    N_{l}=S \mathrm{log}_{2}\left ( \left | \mathcal{M}_{l} \right | \right ),\ \forall l \in \left \{ c,1,\cdots,M \right \},
\end{equation}
\begin{equation}
    r_{l}= \frac{\left \lceil N_{l}\mathrm{min}\left ( \frac{\overline{R}_{l}}{\mathrm{log}_{2}\left | \mathcal{M}_{l} \right |},\beta \right )   \right \rceil}{N_{l}},\ 
    \forall l \in \left \{ c,1,\cdots,M \right \}.
\end{equation}


\subsection{Equalizer}
For each user $k \in \mathcal{K}$, minimum mean square error (MMSE) equalizers are used to detect the common and private streams.
The common stream's equalizer $g_{c,k}^{MMSE}$
is calculated by minimising the MSE 
$\varepsilon _{c,k}=\mathbb{E}\big \{ \left | g_{c,k}y_{k}-{s}_{c} \right | ^{2} \big\}=\left | g_{c,k}\right |^{2} T_{c,k}-2\mathcal{R}\left \{ g_{c,k}\mathbf{h}_{k}^{H} \mathbf{p}_{c} \right \}+1$, where $T_{c,k}=\left|\mathbf{h}_{k}^{H}\mathbf{p}_{c}  \right |^{2}+\left |\mathbf{h}_{k}^{H}\mathbf{p}_{\mu\left ( k \right )}  \right |^{2}+\sum _{j=1,j\neq \mu\left ( k \right )}^{M}\left |\mathbf{h}_{k}^{H}\mathbf{p}_{j}  \right |^{2}+ 1$.
To minimize the MSEs, we let
$\frac{\partial \varepsilon _{c,k}}{\partial g_{c,k}}=0$ and obtain 
\begin{equation}
    g_{c,k}^{MMSE}=\mathbf{p}_{c}^{H}\mathbf{h}_{k}T_{c,k}^{-1} =\frac{\mathbf{p}_{c}^{H}\mathbf{h}_{k}}{\left|\mathbf{h}_{k}^{H}\mathbf{p}_{c}  \right |^{2}+\sum _{j=1}^{M}\left |\mathbf{h}_{k}^{H}\mathbf{p}_{j}  \right |^{2}+ 1}.
\end{equation}
Then, after the common stream is reconstructed and subtracted, the private stream equalizer $g_{k}^{MMSE}$
is calculated by minimising the MSE 
$\varepsilon _{k}=\mathbb{E}\big\{ \left | g_{k}\left (y_{k} - \mathbf{h}_{k}^{H} \mathbf{p}_{c}s_{c}\right )-{s}_{k} \right | ^{2} \big\}=
\left | g_{k}\right |^{2} T_{k}-2\mathcal{R}\left \{ g_{k}\mathbf{h}_{k}^{H} \mathbf{p}_{\mu\left ( k \right )} \right \}+1$, where
$T_{k}= T_{c,k} - \left |\mathbf{h}_{k}^{H}\mathbf{p}_{c}  \right |^{2}$.
By letting $\frac{\partial \varepsilon _{k}}{\partial g_{k}}=0$, the MMSE equalizers for private streams writes as
\begin{equation}
    g_{k}^{MMSE}=\mathbf{p}_{\mu\left ( k \right )}^{H}\mathbf{h}_{k}T_{k}^{-1}
    =
    \frac{\mathbf{p}_{\mu\left ( k \right )}^{H}\mathbf{h}_{k}}{\sum _{j=1}^{M}\left |\mathbf{h}_{k}^{H}\mathbf{p}_{j}  \right |^{2}+ 1}.
\end{equation}

\subsection{Demodulator and Decoder}
We use the Log-Likelihood Ratio (LLR) method detailed in \cite{9217326, 1378448}. LLR is an efficient demodulator in bit-interleaved coded modulation (BICM) systems, and is calculated from the equalized signal for Soft Decision (SD) decoding of polar codes.
A conventional CRC-aided polar decoder is then employed \cite{7055304}.
From Fig. 1, it should be noted that signal reconstruction is performed at the output of the polar decoder. The reconstruction module is the same as the process at the transmitter to reconstruct a precoded signal for the SIC algorithm.

\begin{figure} 
    \centering
    \includegraphics[width=0.43\textwidth]{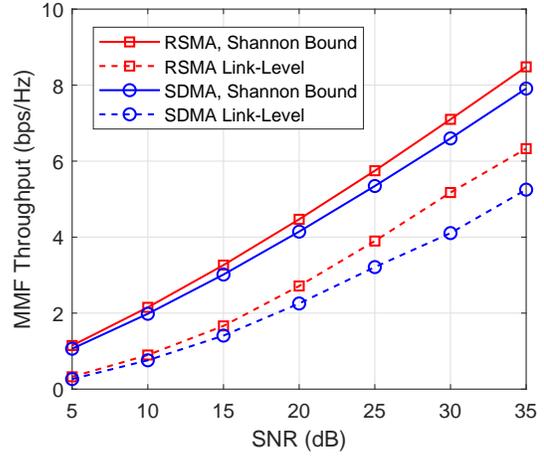}
    \caption{MMF Throughput versus SNR, $\alpha = 0.8$, $N_{t} = 6$ antennas, $K = 6$ users, 2 user per group.}
    \label{fig:Fig2}
\end{figure}

\begin{figure} 
    \centering
    \includegraphics[width=0.43\textwidth]{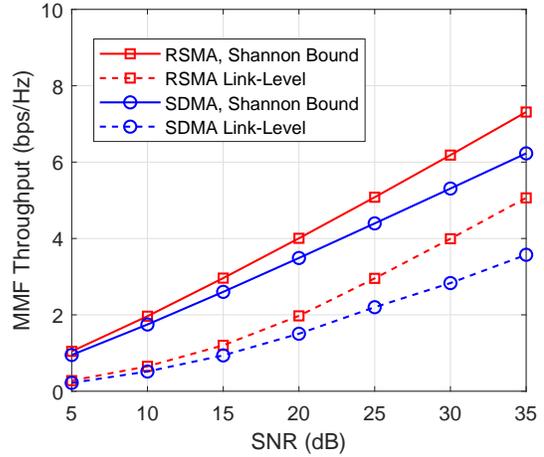}
    \caption{MMF Throughput versus SNR, $\alpha = 0.6$, $N_{t} = 6$ antennas, $K = 6$ users, 2 user per group.}
    \label{fig:Fig3}
\end{figure}

\section{Link-Level Simulation Results}

The superiority of RSMA-based multigroup multicast compared with SDMA under imperfect CSIT has been demonstrated in both terrestrial and satellite communications in \cite{9257433} when considering Gaussian signaling and infinite block length.
SDMA is a special case of RSMA and can be obtained from RSMA by
turning off the common stream and
letting all messages separately encoded into private streams.
The performance gain of RSMA over SDMA is not free as 
the encoding and receiving complexity of RSMA is higher. 
For the one-layer
RSMA in this work, $K + 1$ streams need to
be encoded and $1$ SIC is required at each receiver. For
SDMA, $K$ streams are encoded and no SIC is needed at the receivers.

In this section, 
we demonstrate the performance improvements achieved by RSMA over SDMA for multigroup multicast by LLS results, and compare the obtained throughput levels with the Shannon bounds in \cite{9257433}.
The PHY-layer design follows the architecture described in Fig. 1.
Appropriate modulation schemes and coding rates are selected by the AMC algorithm.

\begin{figure} 
    \centering
    \includegraphics[width=0.43\textwidth]{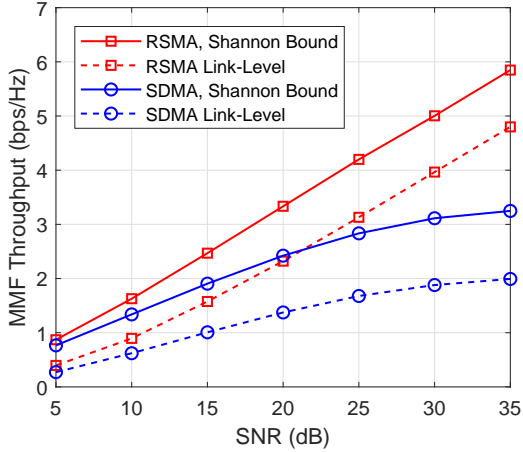}
    \caption{MMF Throughput versus SNR, $\alpha = 0.8$, $N_{t} = 4$ antennas, $K = 6$ users, 2 user per group.}
    \label{fig:Fig4}
\end{figure}

In LLS, we define throughput as the number of bits which can be transmitted correctly at a single channel use.
All MMF throughput levels are obtained by averaging over 100 Monte-Carlo realizations.
The number of channel uses in the $l$-th Monte-Carol realization is denoted by $S^{\left (l  \right )}$.
 $D_{s,k}^{\left ( l \right )}$ denotes the number of successfully recovered information bits by user-$k$ for all $k \in \mathcal{K}$.
 Thus, the MMF throughput can be written as
 \begin{equation}
 \mathrm{MMF \ Throughput}\left [ \mathrm{bps/Hz} \right ] = \frac{ \min_{k \in \mathcal{K}}{\sum _{l}D_{s,k}^{\left ( l \right )}}}{\sum _{l}S^{\left ( l \right )}}.
 \end{equation}
 Without loss of generality, we assume $S^{\left (l  \right )}=256$ for all $l = 1,\cdots,100$ Monte-Carlo realizations.
 The maximum code rate is set as $\beta =0.9$.
 Note that the instantaneous rate may be different from the calculated ARs due to some unexpected loss coming
from the finite length channel coding, finite alphabet modulation and imperfect CSIT.
To ensure correct transmission, energy back-off needs to be performed to the given ARs in order to compensate such losses.
 The energy back-off values for the AMC algorithm are chosen during simulations to maximize MMF throughput while satisfying $\mathrm{BLER}\leq 0.1$ simultaneously for each user.

First, we consider a cellular terrestrial multigroup multicast system with $K = 6$ users equally divided into $M = 3$ multicast groups.
Independent and identically distributed (i.i.d) Rayleigh fading channels are adopted.
When the number of transmit antenna $N_{t}=6$, the system is underloaded. 
Fig. 2 and Fig. 3 respectively show the Shannon bounds and throughput levels achieved by RSMA and SDMA with imperfect CSIT $\alpha = 0.8$ and $\alpha = 0.6$.
It can be clearly observed that RSMA has a significant LLS throughput gain over SDMA in the considered imperfect CSIT scenarios.
The trend of throughput levels is consistent with that of Shannon bounds.
The performance improvements achieved by RSMA over SDMA is demonstrated in the PHY-layer design and LLS platform. 
Moreover, as the CSIT error scaling factor drops from $0.8$ to $0.6$, the CSIT uncertainty increases, and thus leading to lower throughput levels.

\begin{figure} 
    \centering
    \includegraphics[width=0.43\textwidth]{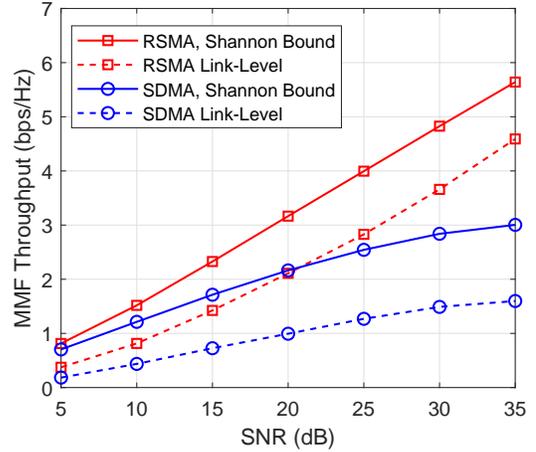}
    \caption{MMF Throughput versus SNR, $\alpha = 0.6$, $N_{t} = 4$ antennas,  $K = 6$ users, 2 user per group.}
    \label{fig:Fig5}
\end{figure}

Fig. 4 and Fig. 5 depict the Shannon bounds and throughput levels when the number of transmit antenna $N_{t}$ is $4$.
 Now the system becomes overloaded, 
 all multiplexing gains of SDMA are sacrificed and collapse to 0 \cite{9257433}. 
 The curve of SDMA Shannon bound gradually saturates as SNR grows. Therefore, the rate gain of RSMA over SDMA is more obvious.
 By LLS, the MMF throughput levels of both RSMA and SDMA follow the trend of Shannon bounds with comparable gaps.
 The throughput of RSMA outperforms SDMA significantly in the presence of considered imperfect CSIT $\alpha = 0.8$ and $\alpha = 0.6$.

\begin{figure} 
    \centering
    \includegraphics[width=0.43\textwidth]{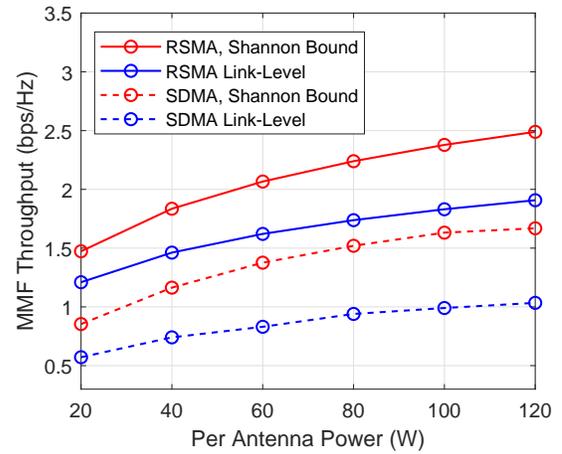}
    \caption{MMF Throughput versus per-antenna available power, $\alpha = 0.8$, $N_{t} = 7$ antennas, $K = 14$ users, 2 user per group.}
    \label{fig:Fig6}
\end{figure}

Then, we consider the same multibeam satellite system as in \cite{9257433} where a geostationary orbit (GEO) satellite equipped with $N_{t} = 7$ antennas serves $K = 14$ single-antenna users simultaneously.
Single feed per beam (SFPB) architecture is used such that only one feed is required to generate one beam (i.e., $N_{t}$ = $M$).
Hence,
$\rho = \frac{K}{M} =2$ users are served simultaneously by each beam, and the system still follows multigroup multicast transmission.
The multibeam satellite channel model is described in \cite{9257433} with the free space loss, radiation pattern and atmospheric fading taken into account.
Fig. 6 illustrates the Shannon bounds and throughput levels achieved by RSMA and SDMA versus an increasing per-antenna transmit power budget under imperfect CSIT $\alpha = 0.8$
We can still observe the matching trends of the Shannon bounds and throughput curves in this satellite setup.
The advantage of using RSMA in multibeam satellite conmmunications over conventional SDMA is demonstrated by LLS even though there are apparent gaps between the throughput levels and Shannon bounds.

\section{Conclusion}

In this work, we evaluate the performance of RSMA for multigroup multicast by PHY layer design and link-level simulations.
The RSMA transmitter and receiver architecture and LLS platform are designed by considering finite length polar coding, finite alphabet modulation, adaptive modulation and coding (AMC) algorithm, and SIC  receivers, etc.
According to numerical link-level results in the considered imperfect CSIT scenarios,
significant throughput gain achieved by RSMA over SDMA for multigroup multicast is demonstrated in a wide range of setups including underloaded/overloaded cellular systems and multibeam satellite systems.
We can conclude that RSMA is a very promising multiple access strategy for practical implementation
to tackle the challenges of modern communication systems in numerous application areas.

\bibliographystyle{IEEEtran}
\bibliography{ref}

\end{document}